\documentclass[%
 reprint,
 bibnote,
 superscriptaddress,
 amsmath,amssymb,
 aps,
 floatfix,
]{revtex4-2}

\usepackage[version=3]{mhchem} 
\usepackage{textcomp}
\usepackage{tikz}
\usepackage{graphicx}
\usepackage{dcolumn}
\usepackage{bm}

\begin{document}

\preprint{APS/123-QED}

\title{Hybrid Epsilon-Near-Zero Modes of Photonic Gap Antennas}

\author{Ashutosh Patri}
\affiliation{Department of Electrical Engineering, Polytechnique Montr\'eal, Montr\'eal, Qu\'ebec H3T1J4, Canada}
\author{K\'{e}vin G. Cogn\'{e}e}
\email{kevin.cognee@gmail.com}
\affiliation{Center for Discovery and Innovation, City College of New York, New York 10031, USA}
\author{David M. Myers}
\affiliation{Department of Engineering Physics, Polytechnique Montr\'eal, Montr\'eal, Qu\'ebec H3T1J4, Canada}
\author{Louis Haeberl\'{e}}
\affiliation{Department of Engineering Physics, Polytechnique Montr\'eal, Montr\'eal, Qu\'ebec H3T1J4, Canada}
\author{Vinod Menon}
\affiliation{Center for Discovery and Innovation, City College of New York, New York 10031, USA}
\author{St\'{e}phane K\'{e}na-Cohen}
\email{s.kena-cohen@polymtl.ca}
\affiliation{Department of Engineering Physics, Polytechnique Montr\'eal, Montr\'eal, Qu\'ebec H3T1J4, Canada}

\begin{abstract}
We demonstrate that in photonic gap antennas composed of an epsilon-near-zero (ENZ) layer embedded within a high-index dielectric, hybrid modes emerge from the strong coupling between the ENZ thin film and the photonic modes of the dielectric antenna. These hybrid modes show giant electric field enhancements, large enhancements of the far-field spontaneous emission rate and a unidirectional radiation response. We analyze both parent and hybrid modes using quasinormal mode theory and find that the hybridization can be well understood using a coupled oscillator model. Under plane wave illumination, hybrid ENZ antennas can concentrate light with an electric field amplitude $\sim$100 times higher than that of the incident wave, which places them on par with the best plasmonic antennas. In addition, the far-field spontaneous emission rate of a dipole embedded at the antenna hotspot reaches up to $\sim$2300 that in free space, with nearly perfect unidirectional emission.
\end{abstract}

\maketitle

Resonances in naturally occurring materials or artificial metamaterials can lead to values of the real part of the dielectric permittivity $\epsilon_{\text{r}} \approx 0$~\cite{fox2001optical,caloz2005electromagnetic}. This leads to unusual optical phenomena, such as infinite phase velocity, light bending and squeezing without reflection, and vanishing local density of optical states (LDOS)~\cite{edwards2008experimental,silveirinha2006tunneling,liberal2017zero}. In particular, close to the epsilon-near-zero (ENZ) frequency, thin films can support two modes with a nearly flat dispersion relation--- a radiative (leaky) mode above the light-line, often referred to as the Ferrell-Berreman mode, and a non-radiative mode below the light-line, sometimes referred to as an ENZ mode~\cite{vassant2012berreman,campione2015theory}. The field enhancement, deep subwavelength confinement and slow light characteristics of these modes have attracted significant attention over the past decade~\cite{vassant2012epsilon,campione2015epsilon}. In contrast to bulk ENZ materials, ENZ thin films are not characterized by a vanishing LDOS and have the potential to both modify and improve the  emission of nearby emitters~\cite{vassant2012berreman,lobet2020fundamental}. These peculiar properties of ENZ films have brought about new possibilities for control of light emission and nonlinear phenomena~\cite{caligiuri2018planar,wen2018doubly,reshef2019nonlinear,passler2019second,bruno2020negative}.

Optical antennas based on subwavelength resonant structures have been at the core of efforts in near-field optics to control light emission and reception~\cite{agio2013optical,novotny2011antennas,biagioni2012nanoantennas}. Recent efforts aimed at combining antennas with ENZ films have shown that this allows on to take advantage of the strong ENZ confinement, while also coupling efficiently to radiation fields. Antennas can serve as impedance matching elements between an ENZ mode and free space~\cite{biagioni2012nanoantennas}. In this regard, most previous studies have focused on metallic antennas placed on extended ENZ thin films~\cite{jun2013epsilon, campione2016near,schulz2016optical,kim2016role, alam2018large,hendrickson2018coupling}. 

In this Letter, we propose \emph{dielectric} antennas, within which a thin ENZ layer has been embedded. Our ENZ Photonic Gap Antennas (PGAs) rely on the strong hybridization between the photonic modes supported by the antenna and the bulk plasmon resonance supported by the ENZ layer. Unlike  previous  demonstrations with metallic antennas and planar ENZ layers, the ENZ films now form an integral part of the antennas. The resulting hybrid modes show giant electric field and spontaneous emission rate (SER) enhancements. In addition, we demonstrate that the gap induced asymmetry  in  the  modes  of  the  PGA  allows  for  a strongly directional  response  in  both  reception  and  reciprocally, emission from embedded emitters. Notably, the PGAs do not have stringent nanolithography constraints and the absence of extrinsic metals is advantageous in many applications such as those where heating must be minimized or those requiring CMOS compatibility~\cite{albella2013low,kuznetsov2016optically,decker2016resonant,bidault2019dielectric}. 

\begin{figure}
\includegraphics[width=8.5cm]{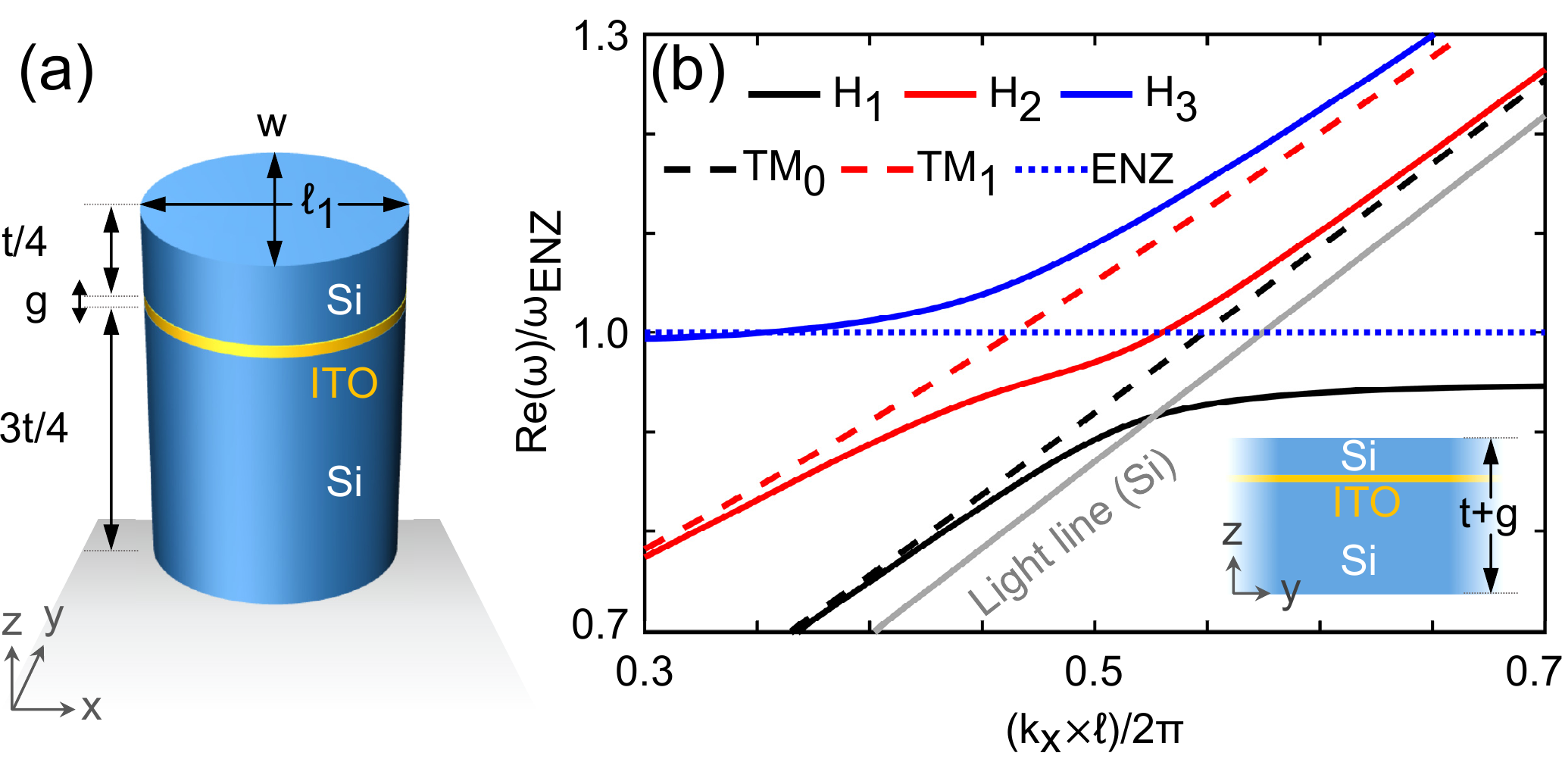}
\caption{\label{fig:figure1} Photonic Gap Antenna (PGA) design. (a) Perspective view of a PGA with design parameters (in nm) $t=580$, $g=2$, $\ell_{1}=300$, and $w=240$. (b) Dispersion relation for the hybrid modes, i.e., $\text{H}_1$ (solid black), $\text{H}_2$ (solid red) and $\text{H}_3$ (solid blue) of a one-dimensional asymmetric ENZ-slot-waveguide with the same layered structure as the PGA (shown in the inset). The dashed lines show the parent modes, i.e., $\text{TM}_0$ (dashed black) and $\text{TM}_1$ (dashed red) of the Si slab waveguide without the ENZ layer. The dotted blue line corresponds to the ENZ frequency of ITO (the longitudinal plasmon resonance). The dispersion relation is plotted in terms of normalized real angular frequency Re$(\omega)/\omega_\text{ENZ}$ as a function of the normalized propagation constant $k_x\ell/2\pi$, where $\ell=200$~nm.}
\vspace{-15pt}
\end{figure}

To illustrate the idea, Fig.~\ref{fig:figure1}(a) shows a PGA composed of a 580~nm thick silicon pillar ($\epsilon_{\text{Si}}=12.5$) within which a thin layer of indium tin oxide (ITO), chosen as the ENZ material, has been embedded. Like highly doped semiconductors, ITO and other transparent conductive oxides have the advantage of having a broadly tunable ENZ frequency range, adjustable by changing the doping concentration~\cite{cleary2018optical}.
Here, a Drude model is used to describe the complex dielectric constant of ITO, with $\text{Re}(\epsilon_{\text{ITO}})=0$ at the ENZ frequency $\omega_{\text{ENZ}}\approx243.6$~THz~\cite{NoteX}. 

To understand the resonant modes of the PGA, it is instructive to first examine the modes of the equivalent one-dimensional Si/ITO/Si waveguide. In this structure, which is effectively a slot waveguide, the field enhancement can be related to the difference in the permittivities between the gap material ($\epsilon_{\text{gap}}$) and the slab ($\epsilon_{\text{slab}}$) material due to continuity of the normal displacement fields $E_{\text{gap}}=(\epsilon_{\text{slab}}/\epsilon_{\text{gap}})E_{\text{slab}}$~\cite{sun2007horizontal}. This immediately hints at interesting consequences resulting from the use of an ENZ layer within the gap ($\epsilon_{\text{gap}}\approx 0$). 

The dashed lines in Fig.~\ref{fig:figure1}(b) show the dispersion relation of the two lowest-order transverse magnetic (TM) guided modes supported by the Si waveguide (thickness, $t=580$~nm) in the absence of the ENZ layer. The $\text{TM}_{0}$ mode has an even $E_{z}$ field distribution along its height, with a maximum field intensity at the center. This suggests that the strongest coupling between the $\text{TM}_{0}$ mode and the ENZ layer will be at this position. However, such a waveguide will produce symmetric modes that lack asymmetry in their radiation pattern on the $yz$ plane required for directionality in the PGA. To break the structural symmetry of the waveguide, we choose an off-centered position for the ENZ layer. In this case, the next mode, $\text{TM}_1$, for which the $E_{z}$ component has two maxima of opposite sign, will also interact with the ENZ layer. In particular, as shown in the inset of Fig.~\ref{fig:figure1}(b), placing the ENZ layer at one of the $\text{TM}_1$ maxima (at a height $\sim t/4$ from the top or bottom) breaks structural symmetry while also allowing for coupling of both modes with the ENZ layer.

The dispersion relation of such an asymmetric Si/ITO/Si waveguide is also shown in Fig.~\ref{fig:figure1}(b) using solid lines. It is known that when the ENZ film thickness (here 2~nm) is reduced below the electric field penetration depth, the longitudinal plasmon resonance supported by the ENZ material at $\omega_{\text{ENZ}}$ acquires transverse character leading to polaritonic ENZ modes~\cite{campione2015theory}. However, the case of an ENZ layer embedded within a finite (along $z$) dielectric differs qualitatively from the extensively-studied cases of free-standing ENZ thin films and ENZ thin films on metals~\cite{campione2015theory,vassant2012berreman,vassant2012epsilon,campione2015epsilon}. Here, we observe clear hybridization between the \emph{guided} modes of the waveguide and the ENZ plasmon resonance as evidenced from the anticrossing behaviour of modes $\text{H}_{1}$, $\text{H}_{2}$ and $\text{H}_{3}$. For relatively low values of the wavenumber ($\sim k_{x} \ell/2\pi < 0.5$) the $\text{H}_{1}$, $\text{H}_{2}$ and $\text{H}_{3}$ modes are $\text{TM}_0$, $\text{TM}_1$ and $\text{ENZ}$-like, respectively. Unlike Ferrell-Berreman modes, however, these modes have no radiative loss since they are guided---only material loss according to their varying ENZ character. Note also that at high wavenumbers $\text{H}_{1}$ $crosses$ the light line of Si and asymptotes below the ENZ frequency with a linear dispersion similar to that of the non-radiative ENZ mode of ITO embedded in infinite Si:  $\omega(k_x)=\omega_{\text{ENZ}}(1-gk_{x}\epsilon_{\text{Si}}/4\epsilon_{\infty}$), where $g$ (2~nm) is the film thickness and $\epsilon_{\infty}$ (3.77) is the high-frequency permittivity limit of the Drude model of ITO. 

In line with our goal to exploit the resonant properties of these hybrid modes in a PGA, we first truncate the Si/ITO/Si waveguide (along $x$) to a two-dimensional structure that satisfies the lowest order Fabry-Perot resonance condition, i.e., $k_{x}\times \ell/2\pi=0.5$, near the anticrossing region ($\sim\omega_{\text{ENZ}}$). For $\ell=200$ nm (along $x$), this condition is satisfied at $\omega_{\text{H1}}=218.2$~THz, $\omega_{\text{H2}}=237.0$~THz, and $\omega_{\text{H3}}=266.2$~THz. The three-dimensional ENZ-slot-waveguide antenna corresponds to a finite width (240~nm) of the waveguide that supports single mode along the $y$-axis. Although all of the important physics are present for truncated-waveguide PGAs, our final PGA design, shown in Fig.~\ref{fig:figure1}(a), has an elliptical cross-section instead. In addition to allowing for simpler fabrication, this tapering of the facets gradually reduces the effective index of the waveguide in a way that further enhances $E_{z}$ in the low-index region~\cite{patri2021photonic}. To maintain the resonance frequencies of the PGA near those of its waveguide counterpart, its length (along $x$) is set to 300~nm. 

\begin{figure}
\includegraphics[width=8.5cm]{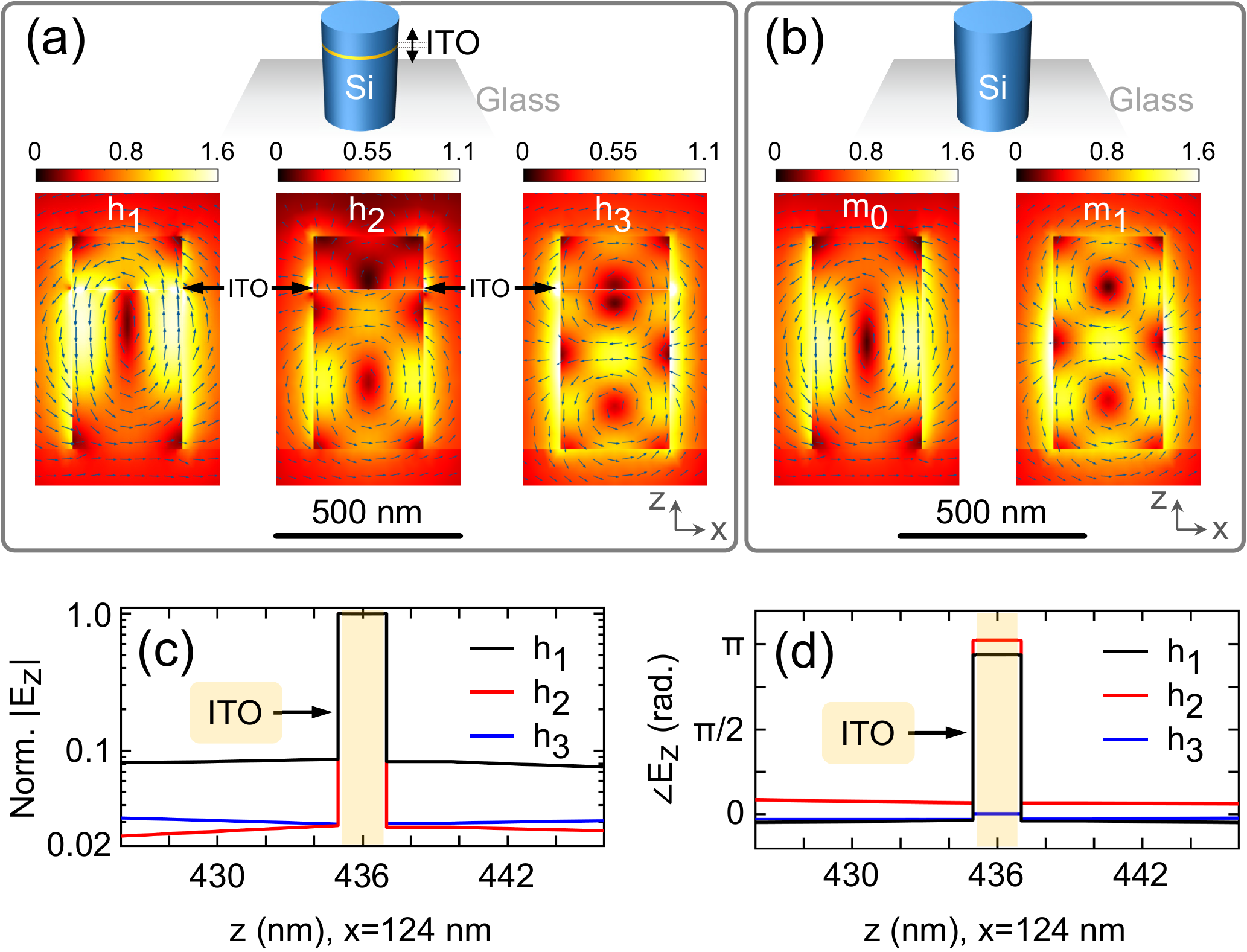}
\caption{\label{fig:figure2} QNM mode profiles. Cuts in the $zx$ plane showing electric fields for the QNMs, (a) $\text{h}_1$, $\text{h}_2$, and $\text{h}_3$ of the PGA, and (b) $\text{m}_0$ and $\text{m}_1$ of the Si antenna, both on a glass substrate, at their corresponding complex frequencies (in THz) $\tilde{\omega}_{\text{h1}}=214.5+12.9$, $\tilde{\omega}_{\text{h2}}=232.7+12.4$, $\tilde{\omega}_{\text{h3}}=255.9+12.4$, $\tilde{\omega}_{\text{m0}}=218.5+15.0$, $\tilde{\omega}_{\text{m1}}=246.4+11.5$. The arrows represent the direction of the real part of the electric field (logarithmic scale) while the color map gives the magnitude of the field. QNM electric fields are expressed in $\times 10^{15}~\text{V}\cdot \text{m}^{-1} \cdot~\text{J}^{-0.5}$. The colorbars are saturated to respective maximum values for a better contrast. The true maximum for QNMs $\text{h}_1$, $\text{h}_2$, $\text{h}_3$, $\text{m}_0$, and $\text{m}_1$ are (in $\times 10^{15}~\text{V}\cdot \text{m}^{-1} \cdot \text{J}^{-0.5}$) $\sim$18.6, 17.2, 16.7, 1.0, and 1.1, respectively. (c) Magnitude and (d) phase of the $z$ component of the normalized electric field along a line, parallel to the $z$-axis, passing through the point of maximum field intensity in the gap $(x,y)=(124,0)$~nm, for $\text{h}_1$ (solid black), $\text{h}_2$ (solid red), and $\text{h}_3$ (solid blue).}
\vspace{-15pt}
\end{figure}

For an accurate description of the localized antenna modes, we use quasinormal mode (QNM) theory~\cite{sauvan2013theory, lalanne2018light}. QNMs are eigensolutions $\{\tilde{\omega}_m,\tilde{\mathbf{E}}_m,\tilde{\mathbf{H}}_m\}$ of linear Maxwell equation for a non-conservative, source-free, and open system, where the initially loaded driving field decays exponentially in time due to the presence of radiation and absorption losses in the system. The electric and magnetic fields ($\tilde{\mathbf{E}}_m,\tilde{\mathbf{H}}_m$) of QNMs are normalized to a unit electromagnetic energy of the system. This allows us to calculate quantities such as the complex mode volume [$\tilde{V}_m \propto 1/\tilde{\mathbf{E}}^2_m(\textbf{r}_0)$] at arbitrary positions $\textbf{r}_0$, which can be directly related to the LDOS~\cite{sauvan2013theory,lalanne2018light}. The decaying QNMs are characterized by complex frequencies $\tilde{\omega}_m=\omega_m+i\gamma_m/2$, where the real and imaginary parts correspond to the resonance frequency and the linewidth of the mode, $m$, respectively. In the results to follow, we analyze PGAs supported by a semi-infinite glass substrate ($\epsilon_{\text{glass}}=2.25$).

In line with the dispersion relation, we identify three hybrid QNMs of the PGA, denoted as $\text{h}_1$, $\text{h}_2$ and $\text{h}_3$. The resulting QNMs field distributions in the $zx$ plane of symmetry are shown in Fig.~\ref{fig:figure2}(a). We can qualitatively compare these hybrid QNMs to the two parent QNMs, denoted as $\text{m}_0$ and $\text{m}_1$ in Fig.~\ref{fig:figure2}(b), of the Si nanopillar without the ITO layer. We observe a close resemblance between $\text{h}_1$ and $\text{m}_0$, and between $\text{h}_3$ and $\text{m}_1$, with the exception of a significant enhancement of the $E_{z}$ components within the ENZ layer for the hybrid QNMs of the PGA. This indicates that $\text{h}_1$ originates from the hybridization of the ENZ resonance with $\text{m}_0$, and $\text{h}_3$ from the hybridization of the ENZ resonance with $\text{m}_1$. In contrast, the field distribution of $\text{h}_2$ has features inherited from both $\text{m}_0$ and $\text{m}_1$, but with strong $E_{z}$ components in the ENZ film and a weak $E_{x}$ component near the top of the PGA as compared to the $E_{x}$ near the substrate---indicating a hybridization of ENZ resonance with both $\text{m}_0$ and $\text{m}_1$. To compare the field intensities of the hybrid modes within ITO to that of Si, at the ITO-Si interfaces, we normalize the field distribution to their maximum, as shown in Fig.~\ref{fig:figure2}(c). The maximum $E_z$ values, as well as the relative difference between the normalized fields within the ITO gap to the adjacent Si, are of similar magnitude for QNMs $\text{h}_2$ and $\text{h}_3$, but with opposite polarity as shown in the phase plot of Fig.~\ref{fig:figure2}(d). This agrees with the fact that the real parts of the permittivity of ITO at the resonant frequencies of $\text{h}_2$ and $\text{h}_3$ are -0.32 and +0.37, respectively.

\begin{figure}
\includegraphics[width=8.5cm]{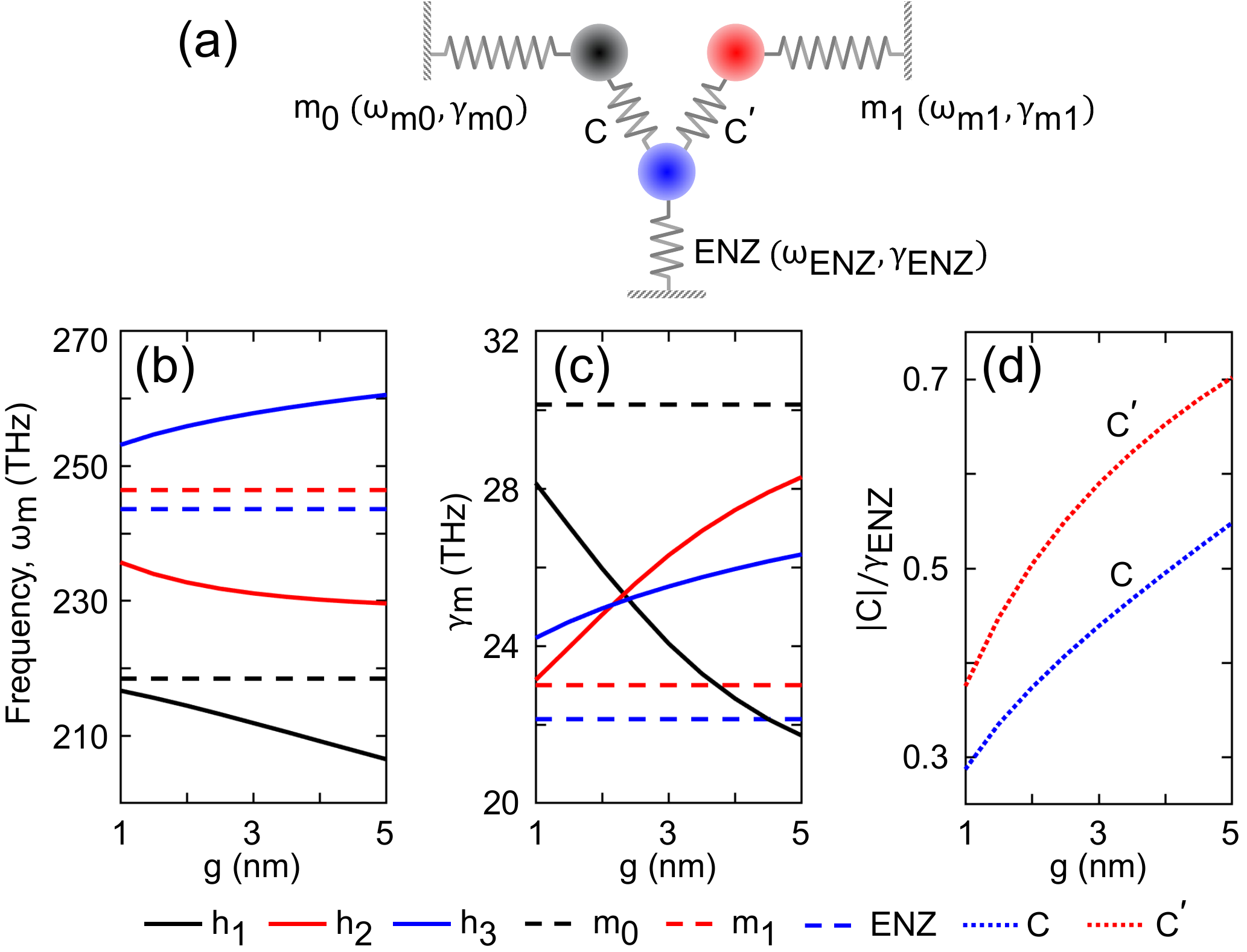}
\caption{\label{fig:figure3} Coupled mode analysis of the hybrid QNMs. (a) Sketch of the coupled oscillator model, where two orthogonal Si antenna modes $\text{m}_0$ (black oscillator) and $\text{m}_1$ (red oscillator) are coupled to ENZ resonance of the ITO layer (blue oscillator) with a coupling coefficient $C$ and $C'$, respectively. (b) Resonance frequency ($\omega_m$) and (c) linewidth ($\gamma_m$) of both hybrid modes (solid lines) and parent modes (dashed lines) for varying thickness of the ITO filled gap layer in the Si antenna. (d) Magnitude of coupling coefficient $C$ (dotted blue) and $C'$ (dotted red), normalized to the ENZ linewidth ($\gamma_{\text{ENZ}}=22.14$~THz), as a function of gap thickness.}
\vspace{-15pt}
\end{figure}

The coupling between the Si antenna and ITO gap layer can be well understood using a coupled oscillator model. The photonic modes of the Si antenna couples to a continuum of ENZ resonance [spatial frequency in the range of $\sim 0.5(2\pi/\ell_{1})$] accumulating in the complex frequency plane around $\tilde{\omega}_{\text{ENZ}}$. For simplicity, we model them as a single mode of complex resonance frequency $\tilde{\omega}_{\text{ENZ}}=\omega_{\text{ENZ}}+i\gamma_{\text{ENZ}}/2$, where $\gamma_{\text{ENZ}}=\gamma_{\text{ITO}}$~\cite{NoteX}. We describe the hybridization between the parent modes with the following $3\times3$ square matrix $\mathbf{A}$, which we assume to be symmetric for simplicity~\cite{tao2020coupling,cognee2020hybridization}.
\begin{equation}
    \mathbf{A}=\left[\begin{matrix}
     \tilde{\omega}_\text{m0} & C & 0   \\
    C&\tilde{\omega}_{\text{ENZ}} & C' \\
    0& C' & \tilde{\omega}_\text{m1}
    \end{matrix}\right].
    \label{eq:model_3x3}
\end{equation}
To account for a mix of coherent and dissipative coupling between oscillators, we consider complex valued coupling coefficients~\cite{tao2020coupling,cognee2020hybridization}. We denote the coefficient between the ENZ resonance and the Si antenna modes, $\text{m}_0$ and $\text{m}_1$, as $C$ and $C'$, respectively. The zeros in the matrix neglect coupling between $\text{m}_0$ and $\text{m}_1$ as they are eigenmodes of the same resonator. Since we know that $\text{h}_1$, $\text{h}_2$ and $\text{h}_3$ are the solution of the eigenproblem  $\mathbf{A}\mathbf{v}_{\pm}=\tilde{\omega}_{\text{h}}\mathbf{v}_{\pm}$, where $\mathbf{v}_{\pm}$ are the eigenvectors, and  $\tilde{\omega}_\text{h}$ are the eigenfrequencies of the hybrid modes, we can numerically fit for the values of the coupling coefficients $C$ and $C'$.

\begin{figure}
\includegraphics[width=8.5cm]{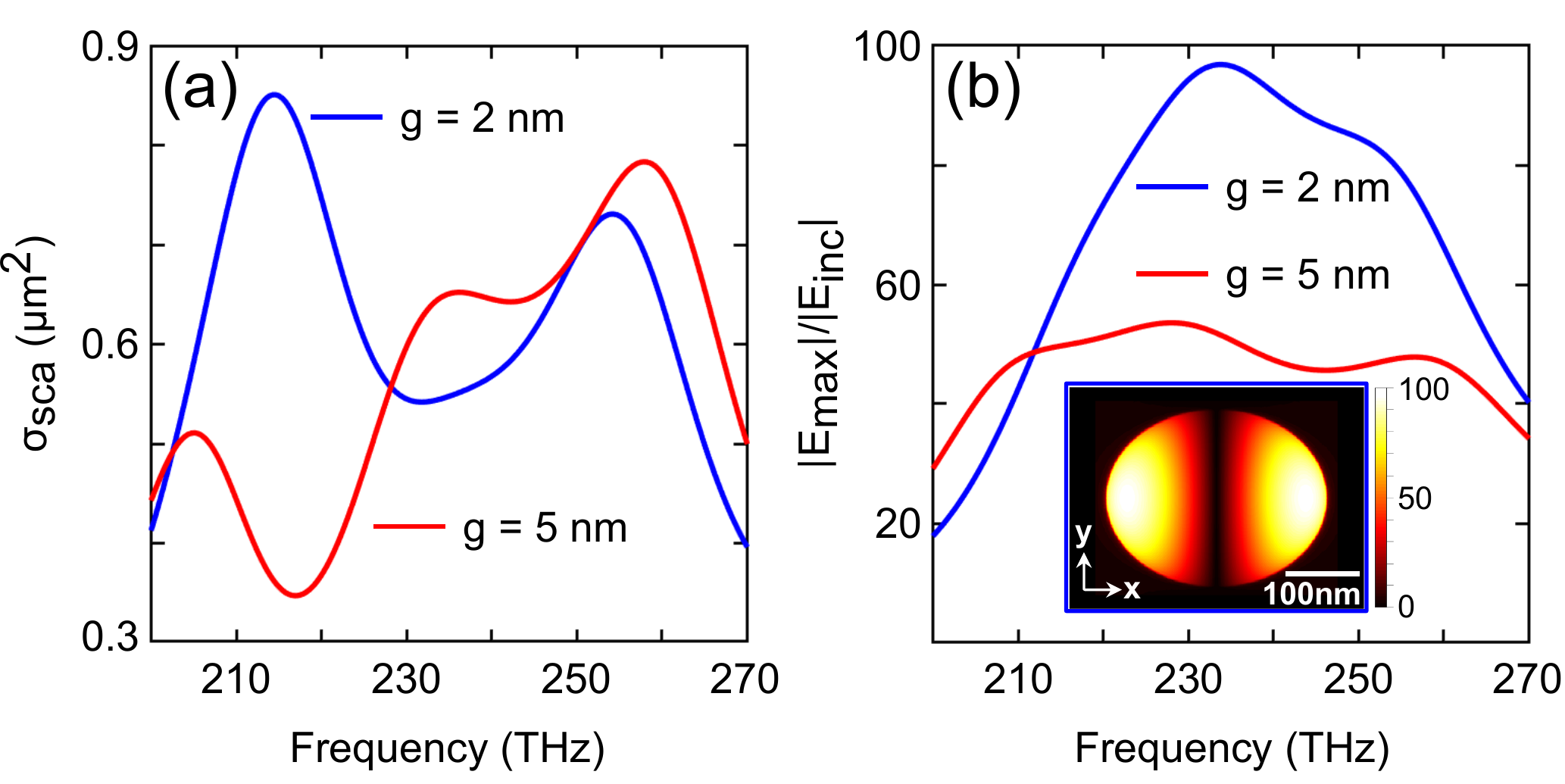}
\caption{\label{fig:figure4} Scattering cross-section and electric field enhancement of the PGAs with 2~nm and 5~nm ITO layers, when illuminated with a $z$ propagating plane wave from the glass-side with its E-field along the $x$-axis. The E-field amplitude is 1 V/m in the glass substrate and this corresponds to an amplitude of $|E_{\text{inc}}|$=1.2 V/m in the air, considering Fresnel reflections at the air-glass interface. (a) Scattering cross-section ($\sigma_{\text{sca}}$) versus frequency, and (b) field enhancement factor $|E_{\text{max}}|/|E_\text{inc}|$ as a function of  frequency. The inset figure shows the distribution of $|E_{\text{max}}|/|E_\text{inc}|$ at $\omega_{\text{h2}}=232.7$~THz for the PGA with a 2~nm gap, on a plane parallel to the $xy$ plane and passing through the center of the gap layer.}
\vspace{-15pt}
\end{figure}

This model allows us to understand the resonant behavior of the PGA as a function of gap thickness. In Figs.~\ref{fig:figure3}(b) and (c), we plot the resonance frequencies and the linewidths, respectively, for both hybrid and parent QNMs. In Fig.~\ref{fig:figure3}(d), we plot the magnitude of fitted coupling coefficients. We observe that the magnitude of $C'$ is bigger than the detuning between ENZ and $\text{m}_1$ indicating strong coupling between the modes. This is in contrast to the nature of coupling between ENZ and $\text{m}_0$, where the detuning is bigger than the magnitude of $C$. We note that as the gap thickness increases, the coupling of ENZ to $\text{m}_0$ increases. This tends to blue-shift $\text{h}_{2}$ and counteract the red-shift originating from the hybridization between ENZ and $\text{m}_{1}$. As shown in Fig.~\ref{fig:figure3}(b), we therefore observe a relatively slow change in the resonance frequency of $\text{h}_{2}$ for increasing gap thickness.  Owing to the higher ($>$ $30\%$) radiation loss rate of $\text{m}_0$ than that of $\text{m}_1$, this increased contribution of $\text{m}_0$ in $\text{h}_{2}$ also increases its radiation loss rate in comparison to $\text{h}_{3}$. In the present design, the coupling strength between the Si antenna and the ITO gap can also be tuned by changing the position of the gap along the height of the antenna to change the overlap between the antenna field and the ENZ layer~\cite{patri2021photonic}.

In practice, the QNMs of any antenna are initially loaded with electromagnetic energy through coupling to a source and then, they decay through coupling to several radiation fields and to material losses. Therefore, we investigate the properties of the PGA when excited by an incident plane wave or a dipole emitter. In the case of plane wave excitation, we evaluate two important properties of the antenna: the coupling between the antenna and the radiation fields as a function of frequency---the scattering cross-section, and its abilities in local electric field enhancement. Whereas, in the case of a dipole emitter, we measure its abilities in spontaneous emission rate (SER) enhancement, and the coupling between the antenna and the radiation fields as a function of direction angles---the directivity pattern.

We begin with the plane wave excitation. Near the substrate, the QNM profiles of the PGA show a relatively strong $E_{x}$ component of even parity. This suggests that an excitation from the bottom will maximize the mode overlap, which in turn increases the power received by the antenna. We plot in Fig.~\ref{fig:figure4}(a) the scattering cross-section $\sigma_{\text{sca}}$ of the PGA for two different gap thicknesses (2~nm and 5~nm). The three peaks in the spectra correspond to the resonance frequencies of the QNMs, but are slightly shifted due to interference between them. The total amount of scattered power is proportional to the received power minus the power absorbed within the ITO layer~\cite{patri2021photonic}. This contribution from absorption loss is observed as a decrease (increase) in $\sigma_{\text{sca}}$ as the values of $\text{Im}(\epsilon_{\text{ITO}})$ increases (decreases) at the resonance frequencies of $\text{h}_1$ ($\text{h}_3$) with increasing gap thickness. The position of the peak corresponding to $\text{h}_2$ does not depend as much as the other two peaks on the thickness of the gap. However, the increasing contribution of $\text{m}_0$ in $\text{h}_2$, for thicker gaps, increases the power received by the antenna.

Another important property of the PGA, in its receiving configuration, is its field enhancement capability ($|E_{\text{max}}|/|E_{\text{inc}}|$), where $E_{\text{max}}$ is the maximum electric field in the vicinity of the antenna and $E_{\text{inc}}$ is the electric field of an incident plane wave. This is shown in Fig.~\ref{fig:figure4}(b) as a function of frequency. Higher values of the field enhancement are obtained when the PGA receives more power and stores it in a smaller three-dimensional space for a longer time. We observe a maximum field enhancement of $\sim$100 in the case of 2~nm ITO gap with an almost homogeneous field distribution along the gap height. These maximum values occur at the point of maximum E-field in the QNM profiles. As the gap thickness increases, the electromagnetic energy is redistributed within the ITO layer and leads to a reduction of the field enhancement that we observe when comparing the results obtained for gaps of 2~nm and 5~nm thickness. Note that the vertical asymmetry in the QNMs of the PGA favours significantly higher reception of power when illuminated from the bottom than the top. Therefore, the antenna should be flipped upside down where high field enhancements are required for an incident plane wave from the top~\cite{patri2021photonic}.

\begin{figure}
\includegraphics[width=8.5cm]{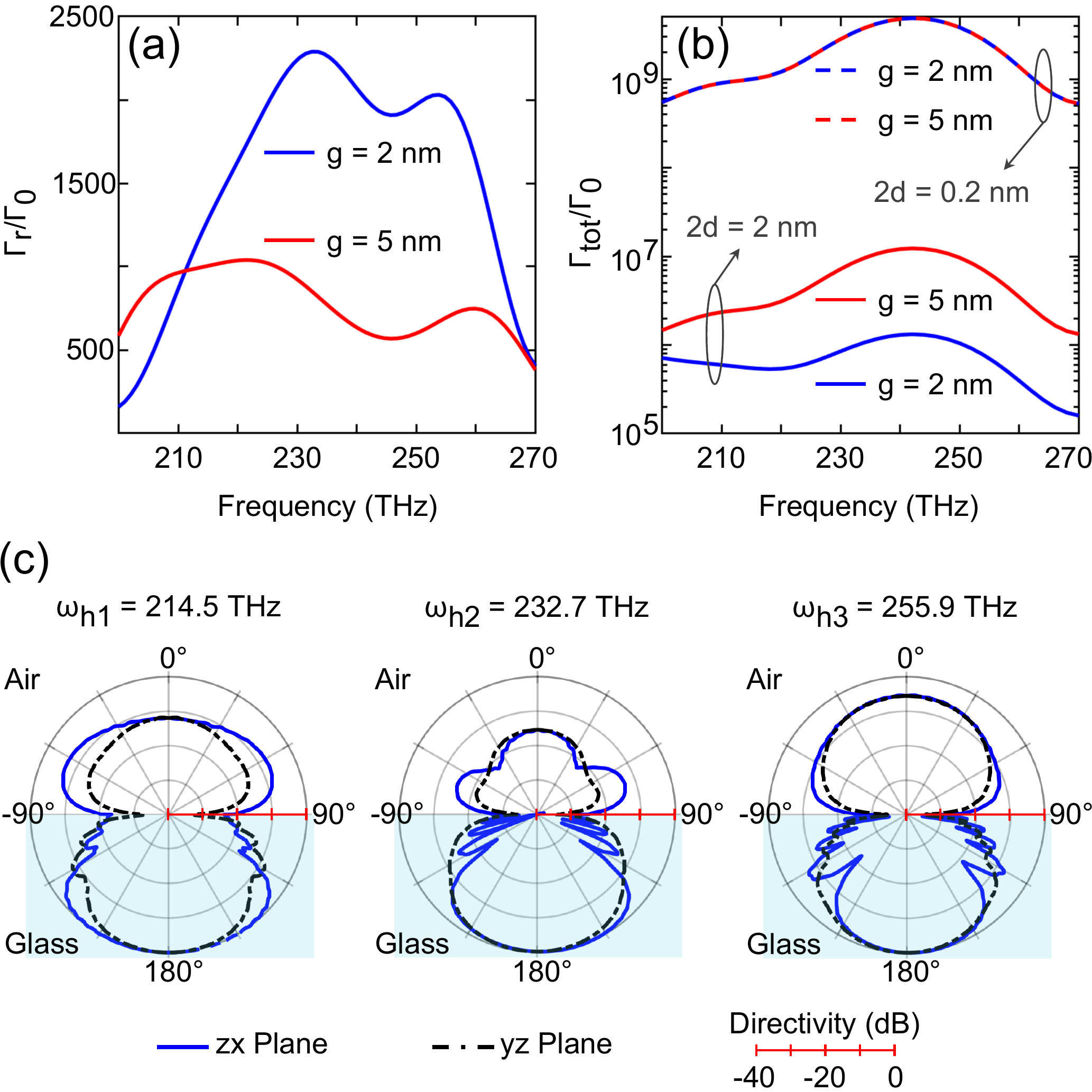}
\caption{\label{fig:figure5} SER enhancement and radiation patterns of an emitter ($z$ oriented) when embedded within 2~nm and 5~nm thick ITO layers of the PGAs. (a) Radiative part of the SER enhancement, $\Gamma_{\text{r}}/\Gamma_{0}$, versus frequency. (b) Total SER enhancement, $\Gamma_{\text{tot}}/\Gamma_{0}$ (logarithmic scale), versus frequency for cut-off conditions $2d=0.2$~nm and $2d=2$~nm. These cut-off conditions are used to approximate the nonlocal response of ITO. (c) Normalized directivity pattern of the PGA with a 2~nm gap, on the $zx$ plane (solid blue) and on the $yz$ plane (dashed black), at the resonance frequencies of the hybrid QNMs calculated from the far-field radiation power of the dipole collected at a distance of 10 \textmu m from the center of the PGA.}
\vspace{-15pt}
\end{figure}

Next, we consider a dipole emitter as the energy source to the hybrid QNMs to study the radiation properties of the antenna in the transmission configuration. In the vicinity of PGA, the dipole radiates through coupling to the available QNMs of the structure. To maximize the coupling strength with the hybrid QNMs, we place the emitter at the maximum E-field of the QNM profiles (the position of maximum LDOS) with its dipole moment oriented along the $z$-axis~\cite{patri2021photonic,sauvan2013theory,lalanne2018light}. In Fig.~\ref{fig:figure5}(a), we plot the radiative part of the SER enhancement of the dipole ($\Gamma_{\text{r}}/\Gamma_0$) as a ratio between the total radiated power collected in the far-field when the dipole is within the PGA ($\propto\Gamma_{\text{r}}$) and when in the free-space ($\propto\Gamma_0$). Similar to the field-enhancement spectra, we observe three peaks in the radiative SER enhancement spectra. An enhancement in the range of $\sim$2000 for the case of 2~nm gap thickness indicates a huge increase in the LDOS and is consistent with the mode volume values of $10^{-3} \lambda_{0}^3$ obtained from our previous QNM analysis. As expected, the radiative SER of the emitter reduces for the case of 5~nm thick gap due to the reduced coupling strength between the emitter and the hybrid antenna modes. At the resonance frequencies of QNMs, the emission rate is predominately dictated by the LDOS of the corresponding QNM. However, at frequencies intermediate of the resonance frequencies, the SER values depend on the combined contribution of QNMs owing to their broad linewidths.  

Along with the three radiative QNMs of the PGA, the emitter also couples to a continuum of non-radiative plasmon modes supported by the ITO layer. This increases the non-radiative part of SER ($\Gamma_{\text{nr}}$) in the system significantly. Given that the emitter is within the ENZ layer, the non-radiative loss will diverge in the absence of a cut-off. In practice, the choice of cut-off depends on the microscopic environment around the emitter. In Fig.~\ref{fig:figure5}(b) the total SER ($\Gamma_{\text{tot}}=\Gamma_{\text{r}}+\Gamma_{\text{nr}}$) enhancement is shown when the emitter is embedded within the 2~nm and 5~nm ITO gaps for two different cut-offs. These cut-off conditions relates to the upper bound of wavevectors that can be excited by the emitter, limited by the nonlocal response of ITO based on momentum conservation~\cite{ford1984electromagnetic}. The first cut-off condition of 2$d$=0.2~nm corresponds to the largest possible wavevector (at $\omega_{\text{ENZ}}$) supported by the conduction band of ITO $k_s=(2m\omega_{\text{ENZ}}/\hbar+k^2_f)^{1/2}+k_f=1/d$, where $k_f$ is the Fermi wavevector and $m$ is the effective mass of the electron (0.35$m_{\text{e}}$). A second, more realistic cut-off of 2$d$=2~nm sets the upper limit for the wavevector at $k_s=1 ~\text{nm}^{-1}$. In our full-wave simulations, we calculate the non-radiative part of SER from the power absorbed in the ITO layer, while excluding the material losses in a cuboid of size 2$d$ surrounding the emitter at its center. This eliminates any possible excitation of larger wavevectors beyond the respective cut-off limits. The peaks observed at $\sim$243~THz and $\sim$210~THz of Fig.~\ref{fig:figure5}(b) can be ascribed to excitation of bulk plasmons and the h$_1$ mode, respectively.

We now show the radiation patterns of the PGA in Fig.~\ref{fig:figure5}(c) at the resonant frequencies of the QNMs. We observe radiation predominantly along the -$z$ direction for all three resonance frequencies. This is primarily because of the asymmetric gap position that leads to vertically asymmetric field distribution of the QNMs, while the effect of substrate is negligible (see Supplementary Fig.~S1). The horizontal asymmetry in the $zx$ plane of radiation containing the electric field vector (E-plane) is due to the asymmetric placement of the emitter along the $x$-axis, which disappears in the $yz$ plane of radiation (H-plane). We observe higher directionality (along -$z$) at 232.7~THz as compared to the other two frequencies. This stems from the strong vertical asymmetry in the $\text{h}_2$ mode combined with the superposition of the $\text{h}_1$ and the $\text{h}_3$ mode in the far-field. This effect can be further improved by tuning the gap position along the height of the antenna (see Supplementary Fig.~S1). Note that this frequency of highest directionality also coincides with the frequency of maximum radiative spontaneous emission, which is in contrast to the case of PGAs with low-index gaps~\cite{patri2021photonic}.

Although our calculations used the experimentally measured dielectric constant of ITO, the achievable performance will depend strongly on the properties of the ENZ medium. For example, scaling to longer wavelengths allows for tighter field confinement using nm-scale gaps. Moreover, the achievable field enhancement and non-radiative loss depend strongly on the ENZ dissipation rate~\cite{javani2016real,khurgin2017replacing}. For example, reducing loss to $0.1\gamma_{\text{ITO}}$ increases the field enhancement up to $\sim$ 200 (see Supplementary Fig.~S2).

In summary, we have introduced an optical antenna based on the PGA concept, where an ENZ material is placed within the gap of a dielectric pillar. This leads to the formation of hybrid ENZ modes due to the coupling between the ENZ resonance to the photonic modes of the antenna. With these hybrid modes, the antenna inherits and even improves upon the strong field confinement capabilities of the thin ENZ films and the radiation properties of the dielectric antennas. The design we describe shows a near-field enhancement of $\sim$100, and SER enhancement by $\sim$2300 along with an unidirectional emission pattern. In addition to the tunable characteristics of ENZ materials, the simplicity of the antenna structure makes it an attractive platform for many applications.

This work was supported by the Natural Sciences and Engineering Council of Canada Strategic Grant program, the Canada Research Chairs program and the Discovery Grants program. The work at CUNY was supported by NSF QII-TAQS grant \#1936351. The authors would like to acknowledge enlightening discussions with A. I. Fern\'{a}ndez-Dom\'{i}nguez about the role of nonlocal effects.


\bibliographystyle{apsrev4-1}
\bibliography{apssamp}

\begin{thebibliography}{39}
\bibitem [{Not()}]{NoteX}%
  \BibitemOpen
  \href@noop {} {}\bibinfo {note} {The Drude permittivity model for ITO can
  described by
  $\epsilon_{\text{ITO}}(\omega)=\epsilon_{\infty}-\omega_p^2/(\omega^2+i~\gamma_{\text{ITO}}~\omega)$,
  where permittivity at infinite frequency $\epsilon_{\infty}=3.77$, plasma
  frequency $\omega_p=473$~THz, and loss rate
  $\gamma_{\text{ITO}}=0.0468~\omega_p$. This gives us
  $\omega_{\text{ENZ}}=\omega_p/\sqrt{\epsilon_{\infty}}$}\BibitemShut
  {NoStop}%
\end{thebibliography}

\end{document}